\documentclass[pra,twocolumn,showpacs,preprintnumbers,amsmath,amssymb,superscriptaddress]{revtex4-1}
\usepackage{graphicx}
\usepackage{dcolumn}
\usepackage{bm}

\begin{document}
\title{Efficient Two-dimensional Subrecoil Raman Cooling of Atoms in a Tripod Configuration}
\author{Vladimir S. Ivanov}\email{ivvl82@gmail.com}
\affiliation{Turku Centre for Quantum
Physics, Department of Physics and Astronomy, University of Turku,
20014 Turku, Finland} \affiliation{Saint Petersburg State
University of Information Technologies, Mechanics and Optics,
197101 St. Petersburg, Russia}
\author{Yuri V. Rozhdestvensky}
\email{rozd-yu@mail.ru} \affiliation{Saint Petersburg State
University of Information Technologies, Mechanics and Optics,
197101 St. Petersburg, Russia}
\author{Kalle-Antti Suominen}\email{Kalle-Antti.Suominen@utu.fi}
\affiliation{Turku Centre for Quantum Physics, Department of
Physics and Astronomy, University of Turku, 20014 Turku, Finland}
\date{\today}

\begin{abstract}
We present an efficient method for subrecoil cooling of neutral atoms by applying Raman cooling in 2D to a four-level tripod-system. The atoms can be cooled simultaneously in two directions using only three laser beams. We describe the cooling process with a simple model showing that the momentum distribution can be rapidly narrowed to velocity spread down to $0.1v_\text{rec}$, corresponding to effective temperature equal to $0.01T_\text{rec}$. This method opens new possibilities for cooling of neutral atoms.
\end{abstract}


\maketitle

\section{Introduction}

Cold atoms are a multipurpose tool for both fundamental observations and technological innovations. Cooling and trapping of neutral atoms has led to many discoveries such as Bose-Einstein condensation~\cite{Pethick2008}, atom interferometry~\cite{Meystre2001}, atomic nanofabrication~\cite{Meschede2003,Oberthaler2003} and improved atomic clocks~\cite{Wynands2005,Boyd2007}. Evaporative cooling in magnetic traps is an efficient tool for reaching high densities at low temperatures~\cite{Ketterle1996}, necessary for the observation of quantum many-body effects in dilute gases. It has been applied with success also to optically trapped gases~\cite{Barrett2001}, as well as to the cooling of fermionic atoms via sympathetic cooling~\cite{DeMarco1999}. However, for many applications it is not necessary or even desirable to reach quantum degeneracy. The external fields and atomic interactions can be harmful for ultraprecise measurements and atomic clocks, for which low temperatures coupled to low densities but large numbers of atoms are needed for a good signal-to-noise ratio and unperturbed atomic transition frequencies. Evaporative cooling is rather wasteful on atoms, and it requires also a large cross-section of elastic collisions and strongly confining traps with velocity-selective output coupling. For quantum information purposes one has to target single atoms, and then cooling without collisional thermalization is needed. Finally, the current interest in nanomechanics has stirred renewed interest on laser cooling~\cite{Cerrillo2010,Machnes2010}. We present an efficient Raman cooling method, which allows one to reach subrecoil temperatures rapidly with purely optical means in 2D, with a simple pulse setup. The method is applicable to all densities and even single atoms, and can be combined with sympathetic cooling if necessary.

The original Raman cooling idea was presented and verified experimentally by
Kasevich and Chu in 1992 at Stanford for Na atoms in 1D~\cite{Kasevich1992}. The
setup consists of a three-level $\Lambda$-type atom. A Raman pulse (two
contra-propagating beams) moves an atom from one state to another and at the
same time gives it a velocity change of $2v_\text{rec}=2\hbar k/m$. The pulse
duration and detunings are selected so that the two-photon Raman process is
suppressed for atoms with velocity near $v=0$. The central frequency and
duration of the subsequent pulses is adjusted to target different velocity
groups while still avoiding $v=0$. As atoms are optically pumped back to the
original state after each pulse, they eventually accumulate at velocities near
$v=0$. The original work was later extended to 2D and 3D
cooling~\cite{Davidson1994}, without reaching subrecoil temperatures. The 1D
scheme has been used in attempts to reach high 3D phase-space densities in
optical traps~\cite{Lee1996,Lee1998}. The use of a dipole trap allowed
subrecoil 3D cooling by ergodicity of the motion in the inverted pyramidal
trap~\cite{Lee1996} or the simple anisotropy of a 3D harmonic
trap~\cite{Lee1996,Perrin1999} while applying the Raman cooling only in 1D.
Similar work with Cs atoms has been done at ENS,
Paris~\cite{Reichel1994,Reichel1995,Perrin1999}. Whereas the original scheme
used Blackman pulses of finite spectral width, it was actually shown with Levy
flight simulations that for 2D and 3D the square pulses are quite
efficient~\cite{Reichel1995}.

The key issue is that in the original scheme the direct Raman cooling in 2D requires four Raman beam pairs and can be quite cumbersome. The 2D scheme with square pulses was applied to Cs atoms at NIST, Gaithersburg, giving temperatures down to 0.15$T_{\rm rec}$~\cite{Boyer2004}, where $T_{\rm rec}$ is the atomic recoil temperature. This is actually the only demonstrated case of reaching subrecoil temperatures with true 2D Raman cooling. We show that in the tripod system, one can go down in theory by an order of magnitude or even more, which means going down to $\frac{1}{100}T_{\rm rec}$. The scheme can also be used in transversal cooling of atomic beams, for example Ne$^*$~\cite{Theuer1999,Vewinger2003}.

The structure of this presentation is such that we first derive the basic equations for the velocity changes induced by one cycle of Raman pulses in the 2D tripod configuration. We show that in the limit of small effective interaction between ground states one can obtain simple solutions for the equations. Then we extend the treatment to arbitrary interaction strength. Finally, we demonstrate the efficiency of the method by solving the model numerically for an appropriate initial momentum distribution, including both the Raman cycles as well as the optical pumping between them.

\section{2D Raman cooling in a tripod configuration}

\subsection{Basic equations}

\begin{figure}
\includegraphics[width=4.5cm,trim=0 -1cm 0 0]{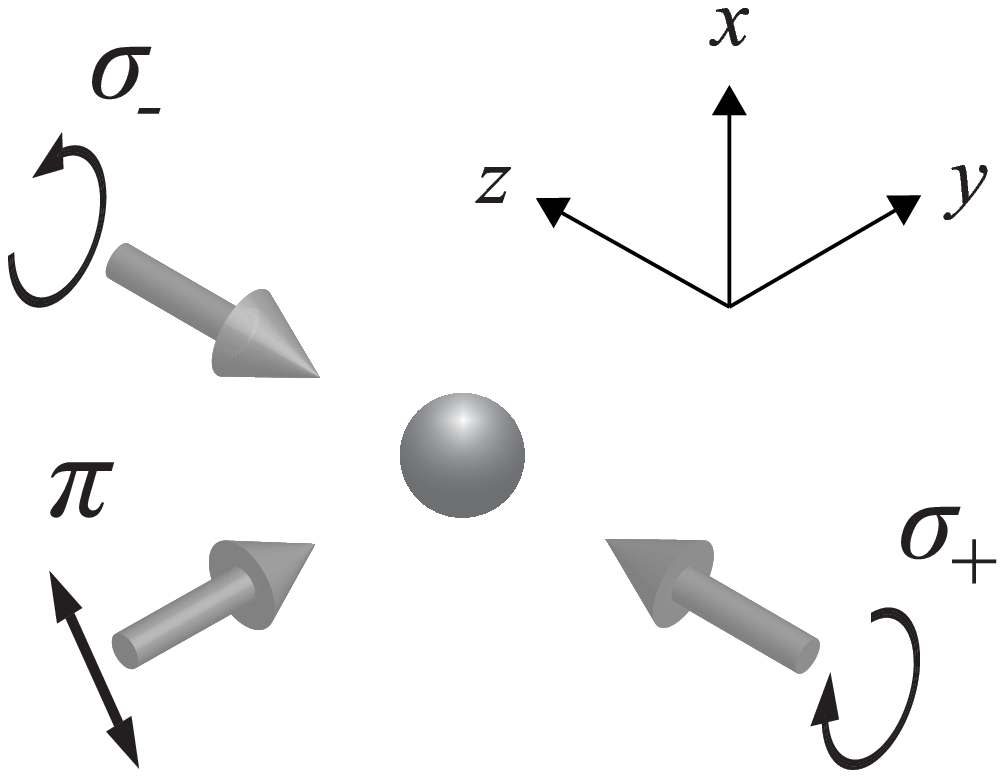}
\includegraphics[width=6.5cm]{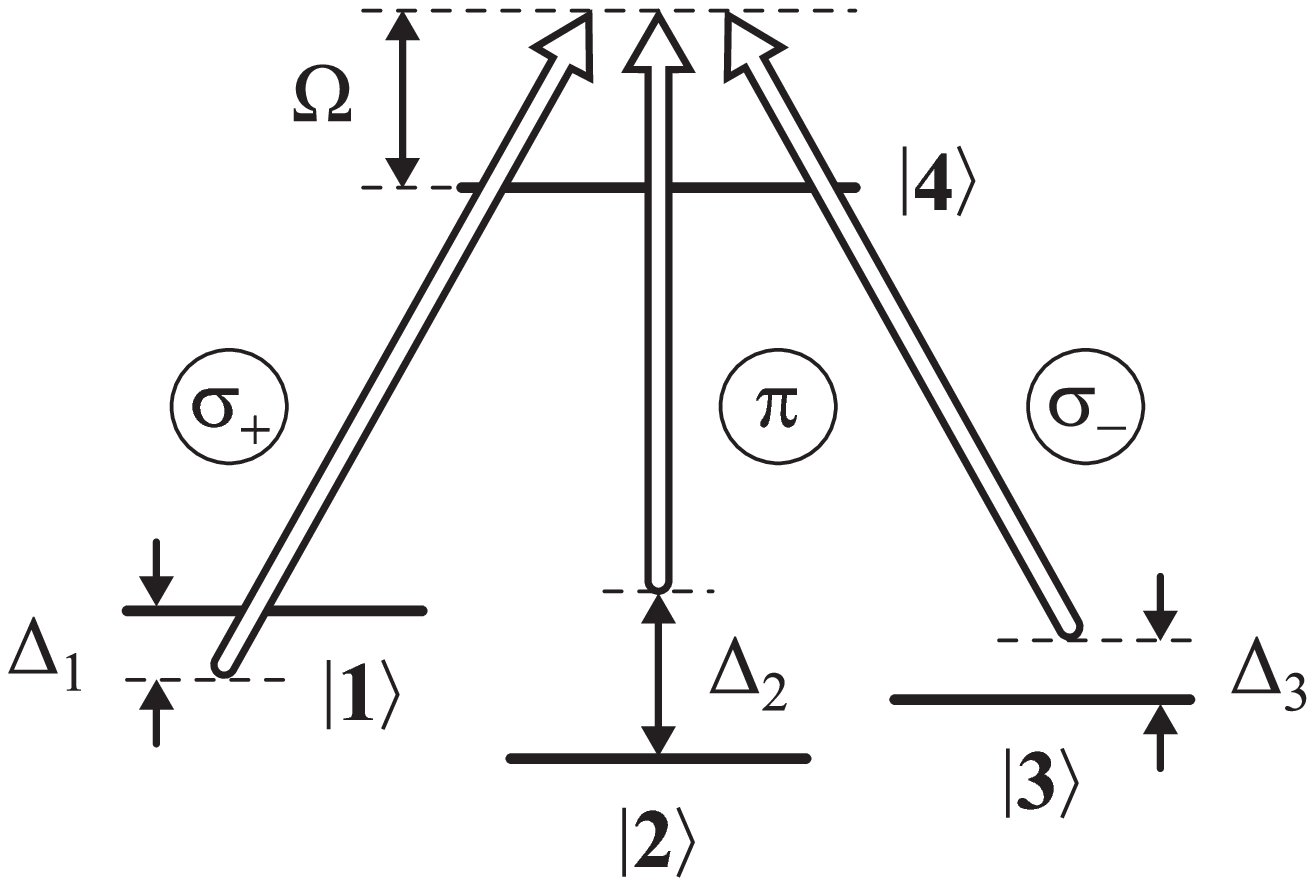}
\caption{\label{scheme} (a) An atomic ensemble interacts with
three running optical waves. (b) The level diagram of a four-level tripod-type atom. The $\sigma_+$ and $\sigma_-$ circularly polarized contra-propagating waves couple transitions $|1\rangle\leftrightarrow |4\rangle$ and $|3\rangle\leftrightarrow |4\rangle$, respectively, and the $\pi$-polarized wave couples the $|2\rangle\leftrightarrow |4\rangle$ transition.}
\end{figure}

Consider an atomic ensemble in the region of interaction with laser configuration that consists of three running optical waves shown in Fig. \ref{scheme}(a). Two contra-propagating $\sigma_{+}$
and $\sigma_{-}$ circularly polarized waves are directed along the $z$ direction, a $\pi$-polarized optical wave propagates along the $y$ direction. The positive frequency part of the electric field can be written as
\begin{equation*}
\vec E^{(+)}(\vec r,t)=\vec E_1e^{-i\omega_1 t+ikz}+\vec
E_2e^{-i\omega_2 t+iky}+\vec E_3e^{-i\omega_3 t-ikz},
\end{equation*}
where $k$ is the wave number same for each optical wave. The first term corresponds to a $\sigma_+$ circularly polarized wave of frequency $\omega_1$ and the third one corresponds to a $\sigma_-$ circularly polarized wave of frequency $\omega_3$. The second term describes a $\pi$-polarized wave of frequency $\omega_2$.

The interaction of a four-level tripod-type atom with the laser
beams is described by the Schr\"odinger equation
\begin{equation}\label{Schr}
i\hbar\frac{\partial\Psi}{\partial
t}+\frac{\hbar^2}{2M}\nabla^2\Psi=\hat H\Psi.
\end{equation}
The atomic Hamiltonian and the probability function in the
rotating-wave approximation (RWA) and in the resonant
approximation are given by
\begin{equation}
\hat H=-\hbar\begin{pmatrix} -\Delta_1&0&0&Ge^{-ikz}\\
0&-\Delta_2&0&Ge^{-iky}\\0&0&-\Delta_3&Ge^{ikz}\\
Ge^{ikz}&Ge^{iky}&Ge^{-ikz}&\Omega\end{pmatrix},
\Psi=\begin{pmatrix}a_1\\a_2\\a_3\\a_4\end{pmatrix},
\end{equation}
where $\Omega$ is the detuning of driving fields from the upper
state, $\Delta_m=\omega_m-\omega_{4m}-\Omega$ ($m=1,2,3$) are
detunings from ground states $|m\rangle$ (see Fig.
\ref{scheme}(b)). Without losing generality we have assumed that
the Rabi frequencies $G_j=|\vec d_{4j}\vec E_j/\hbar|$ ($j=1,2,3$)
equal a real magnitude $G$.

The probability amplitudes in the momentum-space approach are
obtained by using the Fourier transform
\begin{equation*}
a_m(p_z,p_y,t)=\frac{1}{2\pi}\iint\limits_{-\infty}^{\quad\infty}
a_m(z,y,t)e^{-ik(p_zz+p_yy)}dzdy,
\end{equation*}
where $p_z$, $p_y$ are projections of the atomic momentum on the
axes Oz, Oy in units of $\hbar k$. For variables
\begin{equation}\label{subs}
\begin{split}
b_1=a_1(p_z-1,p_y),\quad &b_2=a_2(p_z,p_y-1),
\\b_3=a_3(p_z+1,p_y),\quad &b_4=a_4(p_z,p_y),
\end{split}
\end{equation}
the Schr\"odinger equation (\ref{Schr}) can be written as
\begin{subequations}\label{eq-b}
\begin{align}
i\dot b_1&=\left((p_z-1)^2+p_y^2+\delta_1\right)b_1-gb_4,\\
i\dot b_2&=\left(p_z^2+(p_y-1)^2+\delta_2\right)b_2-gb_4,\\
i\dot b_3&=\left((p_z+1)^2+p_y^2+\delta_3\right)b_3-gb_4,\\
i\dot b_4&=\left(p_z^2+p_y^2-\Delta\right)b_4-gb_1-gb_2-gb_3,
\label{eq-b4}
\end{align}
\end{subequations}
where the dimensionless time, detunings and the Rabi frequency are given by
\begin{equation}\label{scaling}
\tau=\omega_R t,\quad\delta_m=\frac{\Delta_m}{\omega_R},
\quad\Delta=\frac{\Omega}{\omega_R},\quad g=\frac{G}{\omega_R},
\end{equation}
and $\omega_R=\hbar k^2/2M$ is the recoil frequency.

The upper state $|4\rangle$ can be adiabatically eliminated in the case of Raman transitions between the ground states when
$\Delta\gg g$. In this case, we assume that $\dot b_4\approx 0$ and obtain from Eq. (\ref{eq-b4}) expression
\begin{equation}\label{appr}
b_4\approx -\frac{g}{\Delta}(b_1+b_2+b_3).
\end{equation}
Terms $p_z^2$, $p_y^2$ were dropped here as values negligible in comparison with $\Delta$. The substitution of Eq. (\ref{appr}) into Eqs. (\ref{eq-b}) leads to equations
\begin{subequations}\label{3eq}
\begin{align}
i\dot b_1-\left((p_z-1)^2+p_y^2+\delta_1+\alpha\right)b_1&=
 \alpha b_2+\alpha b_3,\label{eq-b1}\\
i\dot b_2-\left(p_z^2+(p_y-1)^2+\delta_2+\alpha\right)b_2&=
 \alpha b_1+\alpha b_3,\label{eq-b2}\\
i\dot b_3-\left((p_z+1)^2+p_y^2+\delta_3+\alpha\right)b_3&=
 \alpha b_1+\alpha b_2\label{eq-b3},
\end{align}
\end{subequations}
where $\alpha=g^2/\Delta$ is the Rabi frequency of the two-photon resonance.

If the atomic population is initially concentrated in state
$|2\rangle$ with corresponding amplitude $b_2^0$, it follows from Eqs. (\ref{3eq}) that during the interaction
\begin{equation}\label{abs}
|b_1|^2+|b_2|^2+|b_3|^2=|b_2^0|^2.
\end{equation}
To solve Eqs.~(\ref{3eq}) one can consider some special cases. These are useful for gaining insight about the parameter choices such as pulse duration that optimize the efficiency of the approach.

\subsection{Short-time interaction}

Let us consider the case of the short-time interaction when
$\alpha\tau\ll 1$. In such a case, variables $b_1,b_3\approx 0$ and therefore $b_1,b_3$ vanish from the right-hand side of Eqs. (\ref{3eq}). After that, Eq.~(\ref{eq-b2}) can be solved separately, and the substitution of $b_2$ into Eqs.~(\ref{eq-b1}) and (\ref{eq-b3}) leads to the solutions
\begin{equation}\label{b1,3}
|b_1|^2\approx |b_2^0|^2\alpha^2\frac{\sin^2 D_1\tau}{D_1^2},\quad
|b_3|^2\approx |b_2^0|^2\alpha^2\frac{\sin^2 D_2\tau}{D_2^2},
\end{equation}
where the denominators are
\begin{equation}\label{denom}
D_1=\frac{1}{2}(\delta_1-\delta_2)-p_z+p_y,\quad
D_2=\frac{1}{2}(\delta_3-\delta_2)+p_z+p_y.
\end{equation}
By substituting Eqs. (\ref{b1,3}) into Eq.~(\ref{abs}), we get
\begin{equation}\label{b2}
|b_2|^2\approx |b_2^0|^2\left(1-\alpha^2\frac{\sin^2
D_1\tau}{D_1^2}-\alpha^2\frac{\sin^2 D_2\tau}{D_2^2}\right).
\end{equation}
Then the atomic populations in ground states, using Eqs.~(\ref{subs}), (\ref{b1,3}) and (\ref{b2}), can be written as
\begin{align}
|a_1(p_z,p_y)|^2 &\approx
|a_2^0(p_z{+}1,p_y{-}1)|^2\alpha^2\dfrac{\sin^2
(D_1-1)\tau}{(D_1-1)^2},\label{a1}
\\
\begin{split} |a_2(p_z,p_y)|^2 &\approx
|a_2^0(p_z,p_y)|^2\left(1-\alpha^2
\frac{\sin^2 (D_1+1)\tau}{(D_1+1)^2}\right.\\
&\left.-\alpha^2\frac{\sin^2 (D_2+1)\tau}{(D_2+1)^2}\right),
\end{split}\label{a2}
\\
|a_3(p_z,p_y)|^2 &\approx
|a_2^0(p_z{-}1,p_y{-}1)|^2\alpha^2\dfrac{\sin^2
(D_2-1)\tau}{(D_2-1)^2}.\label{a3}
\end{align}

\begin{figure}
\includegraphics[width=6cm]{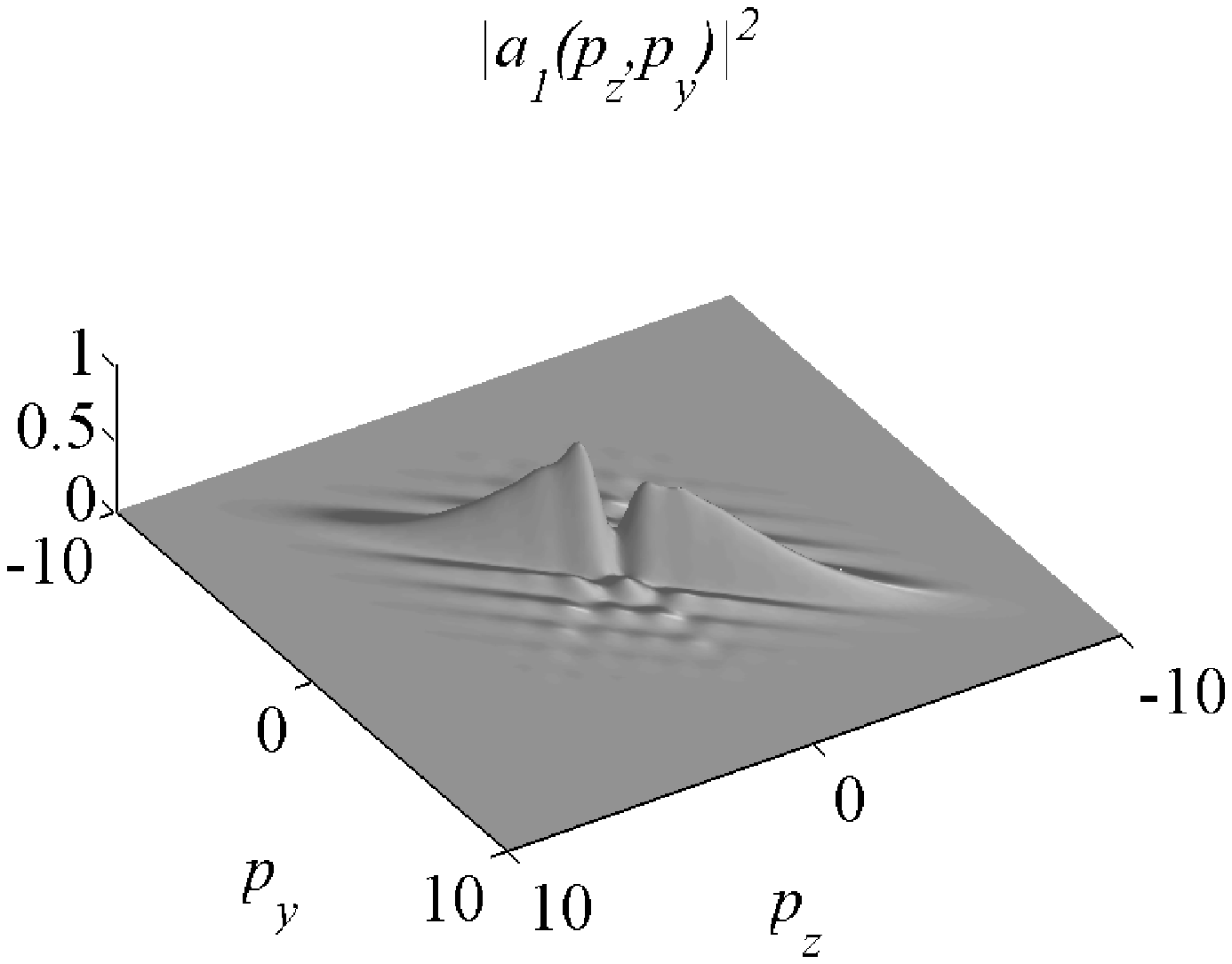}\\(a)\\
\includegraphics[width=6cm]{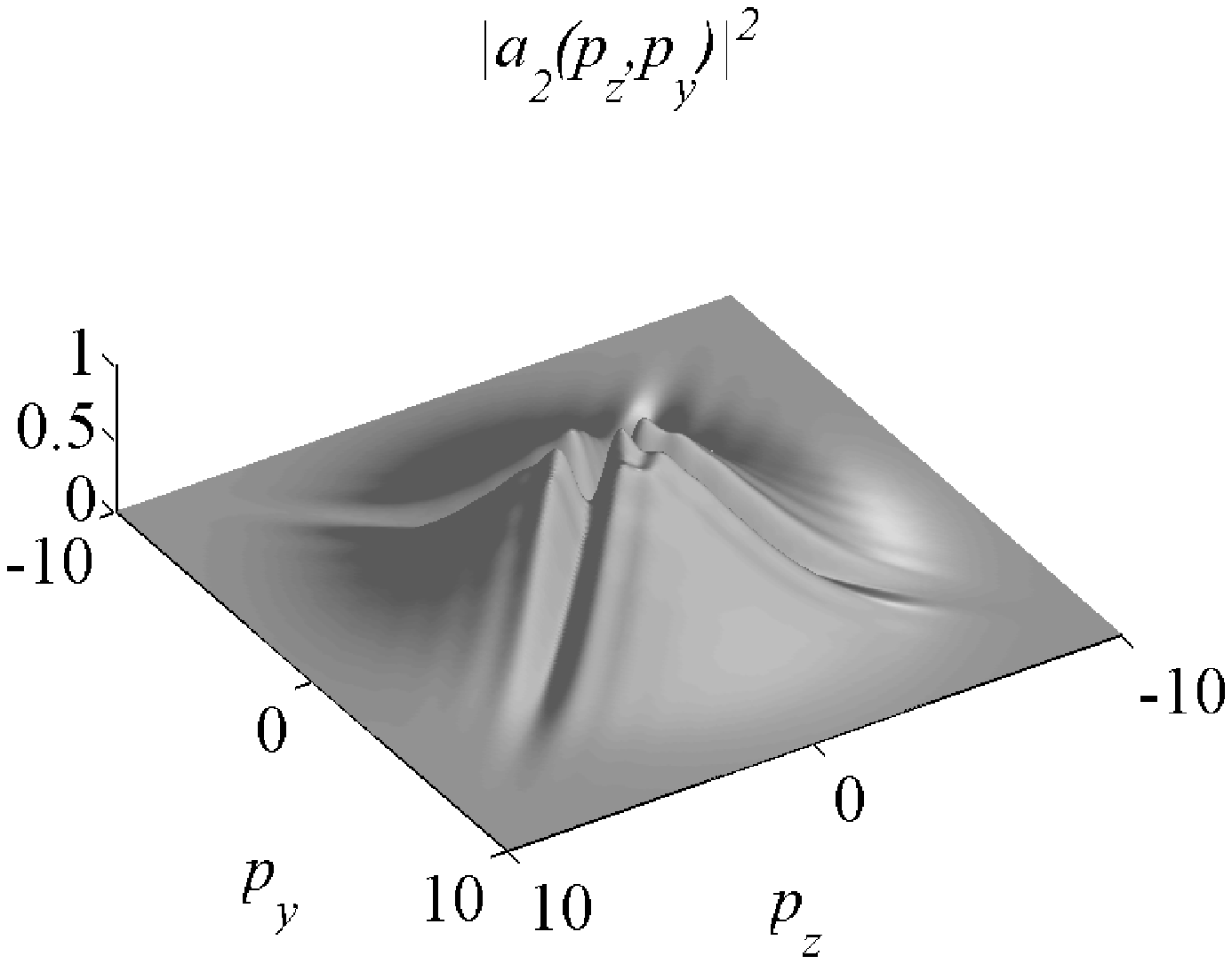}\\(b)\\
\includegraphics[width=6cm]{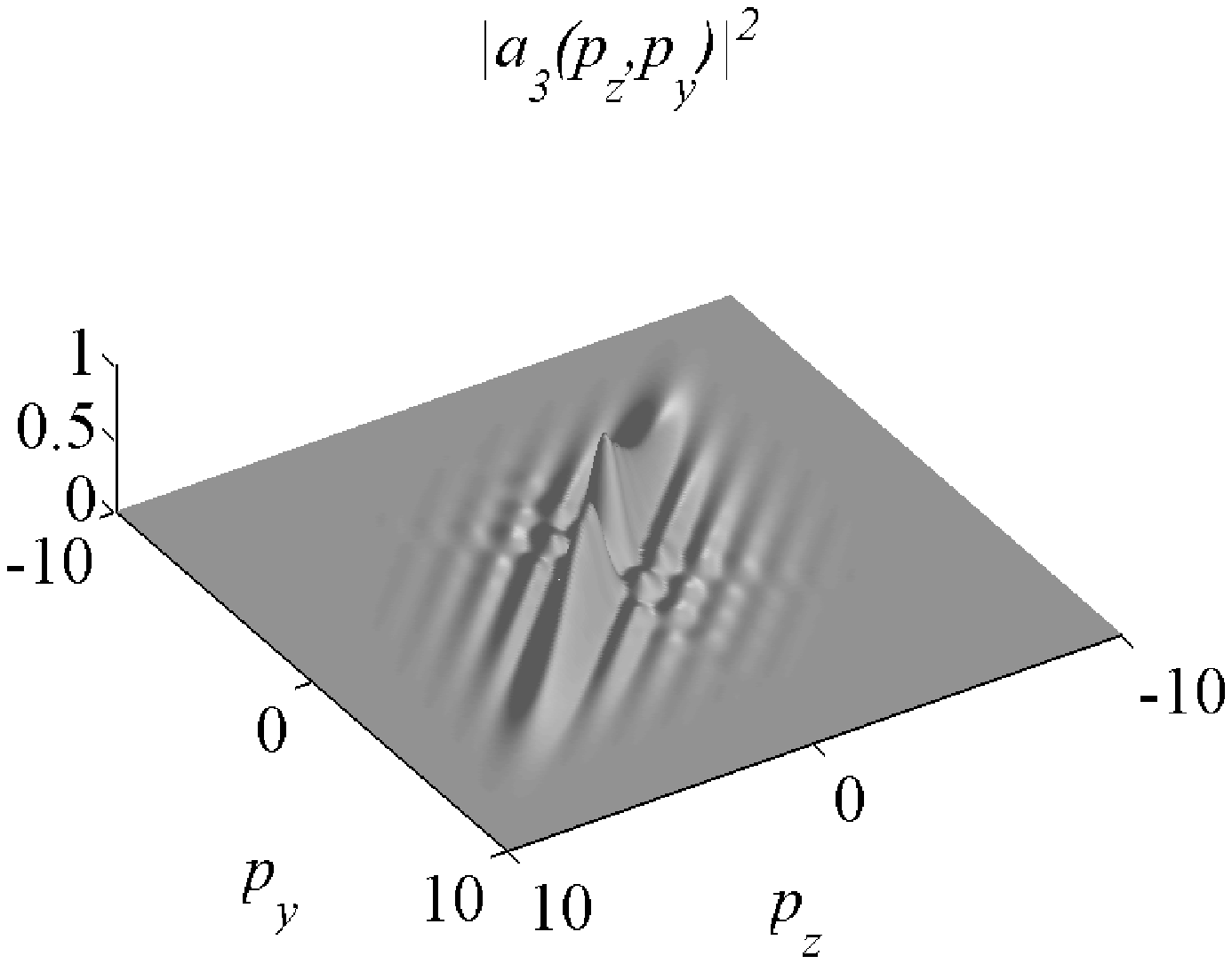}\\(c) \caption{\label{across}
The momentum distribution of atoms originally prepared in state $|2\rangle$ after they interact with a
pulse of Raman beams. (a), (c) The population transferred to states $|1\rangle$, $|3\rangle$ through Raman transitions $|2\rangle\leftrightarrow |1\rangle$ and $|2\rangle\leftrightarrow |3\rangle$, respectively; (b) cross appearing in state $|2\rangle$ due to these transfers. The duration of Raman pulse is $\tau=2$, the Rabi frequency of two-photon resonance is $\alpha=0.5$. The detunings from ground states are $\delta_m=0$ ($m=1,2,3$).}
\end{figure}

Let us consider atoms originally prepared in state $|2\rangle$ after they interact with a pulse of Raman beams. Corresponding approximate forms of the populations in the momentum-space
approach are given by Eqs.~(\ref{a1})-(\ref{a3}). The population transferred to state $|1\rangle$ through Raman
transition $|2\rangle\leftrightarrow |1\rangle$ is shown in Fig. \ref{across}(a), while the population transferred to state $|3\rangle$ through Raman transition $|2\rangle\leftrightarrow |3\rangle$ is shown in Fig. \ref{across}(c). The removed population of state $|2\rangle$ represents a cross as shown in Fig. \ref{across}(b). The width of the cross is defined by the
pulse duration $\tau$. As seen from Eqs. (\ref{denom}) and (\ref{a2}), the largest excitation probability is reached when
\begin{equation*}
\frac{1}{2}(\delta_1-\delta_2)-p_z+p_y+1=0\quad\text{or}\quad
\frac{1}{2}(\delta_3-\delta_2)+p_z+p_y+1=0.
\end{equation*}
The position of the cross can be changed by adjusting the detunings from the ground states, allowing one to excite atoms with any two projections of the atomic momentum. The excitation of atoms through two Raman transitions at once essentially increases the efficiency of cooling in a tripod configuration. Note that for Fig.~\ref{across} we have used $\alpha\tau=1$ but the short-interaction limit gives a good account for the basic aspects of the process as the relevant equations are modified only slightly for increasing $\alpha\tau$ (see Sec.~\ref{arbit}).

Similarly to 1D Raman cooling, an elementary cycle of 2D Raman cooling consists of two steps. In the first step, atoms prepared in state $|2\rangle$ are selectively transferred by a pulse of Raman beams. In order to suppress the Raman transfer of atoms with the momentum projection $p_z,p_y=0$, we can choose the pulse duration
\begin{equation}\label{tau}
\tau=\frac{2\pi}{\delta+2},
\end{equation}
where $\delta_1-\delta_2=\delta_3-\delta_2=\delta$.
In the second step, atoms return to state $|2\rangle$ via optical pumping. The $\pi$-polarized laser is switched off, the $\sigma_+$, $\sigma_-$ lasers are tuned into resonance
with transitions $|1\rangle\leftrightarrow |4\rangle$ and
$|3\rangle\leftrightarrow |4\rangle$, respectively. An atom is excited to state $|4\rangle$ with a gain of momentum along the $z$ direction, so $p'_z=p_z\pm 1$. Then the atom returns to state $|2\rangle$ emitting a photon with momentum $\Delta\vec p$ where $|\Delta\vec p|=1$. Because of momentum conservation, the atomic momentum changes by $-\Delta\vec p$. So, the atomic population in state $|2\rangle$ becomes
\begin{align*}
|a'_2(p_z,p_y)|^2&=|a_2(p_z,p_y)|^2+|a_1(p_z{-}1{+}\Delta
p_z,p_y{+}\Delta p_y)|^2\\&+|a_3(p_z{+}1{+}\Delta
p'_z,p_y{+}\Delta p'_y)|^2.
\end{align*}

Different momentum groups can be excited from state $|2\rangle$ by adjusting the duration $\tau$ and the difference of detunings $\delta$ in consistent with (\ref{tau}). The set of detunings $\delta$ used by us is
\begin{equation}\label{set}
\delta=2^k-2,\quad k=0,1,..,5.
\end{equation}
The position of the cross for large $\tau$ is situated in
immediate proximity to zero projections, which gives as a result that only very cold atoms are left in state $|2\rangle$.

\subsection{Arbitrary Rabi frequency solution}\label{arbit}

To simplify the task, we assume that $\delta_1 = \delta_3 = \delta_2 + \delta$, and consider the only atoms of momentum projection $p_z=0$. These atoms are characterized by the coherence between states $|1\rangle$ and $|3\rangle$, which in turn is derived from the difference of Eqs. (\ref{eq-b1}) and (\ref{eq-b3})
\begin{equation}\label{difb1,3}
i\frac{d}{d\tau}(b_1 - b_3) -
\left(1 + p_y^2 + \delta_1 \right) (b_1 - b_3) = 0.
\end{equation}
Before the Raman pulse starts, the atoms are contained in state $|2\rangle$. Hence $b_1^0, b_3^0 = 0$, and one obtains from Eq. (\ref{difb1,3}) that
\begin{equation}\label{eqvb1,3}
b_1 = b_3.
\end{equation}
After the substitution of Eq. (\ref{eqvb1,3}), the equations (\ref{3eq}) are reduced to a system of two equations
\begin{align}
i\dot b_1 &= \left( 1 + p_y^2 + \delta_1 + 2\alpha \right) b_1
+ \alpha b_2,
\\
i\dot b_2 &= \left( (p_y-1)^2 + \delta_2 + \alpha \right) b_2
+ 2\alpha b_1.
\end{align}
Then the interaction with Raman beams is described by the effective Hamiltonian and probability function
\begin{align}\label{eff}
\hat H_\text{eff} = -\hbar\begin{pmatrix}
0 & \alpha_\text{eff} \\
\alpha_\text{eff} & \delta_\text{eff}
\end{pmatrix},
\Psi_\text{eff} = \begin{pmatrix} -\sqrt{2} b_1 \\ b_2 \end{pmatrix}
e^{i( 1 + p_y^2 + \delta_1 + 2\alpha )\tau},
\end{align}
where $ \delta_\text{eff} = \delta + 2p_y + \alpha $,
$ \alpha_\text{eff} = \sqrt{2}\alpha $.

The Hamiltonian (\ref{eff}) in the dressed-state picture is presented by the eigenstates of the probability amplitudes
\begin{align}\label{eigenst}
\begin{aligned}
\psi_+ = \psi_1 \cos\phi - \psi_2 \sin\phi,
\\
\psi_- = \psi_1 \sin\phi + \psi_2 \cos\phi,
\end{aligned}
\quad
\tan\phi = \frac{
\sqrt{ 4\alpha_\text{eff}^2 + \delta_\text{eff}^2 } - \delta_\text{eff} }
{2\alpha_\text{eff}},
\end{align}
with the corresponding eigenfrequencies
\begin{align*}
\omega_\pm = -\frac{\delta_\text{eff}}{2}
\pm \frac{1}{2}\sqrt{ 4\alpha_\text{eff}^2 + \delta_\text{eff}^2 },
\end{align*}
where $\psi_1$, $\psi_2$ are the probability amplitudes of states in presentation (\ref{eff}). Expression (\ref{eigenst}) gives original $\psi_\pm$ after the substitution of $\psi_1^0 = 0$, $\psi_2^0 = b_2^0$, i.e., the probability amplitudes are written as
\begin{align}\label{psi}
\psi_+ = - b_2^0 \sin\phi \, e^{-i\omega_+ \tau},
\quad
\psi_- = b_2^0 \cos\phi \, e^{-i\omega_- \tau}.
\end{align}
By inverting Eq.~(\ref{eigenst}) we get
\begin{align*}
\psi_1 = \psi_+ \cos\phi + \psi_- \sin\phi,
\quad
\psi_2 = -\psi_+ \sin\phi + \psi_- \cos\phi,
\end{align*}
which gives via Eq.~(\ref{psi}) the probability amplitudes
\begin{align*}
\psi_1 &= \frac{b_2^0}{2} \sin 2\phi
\left( e^{-i\omega_- \tau} - e^{-i\omega_+ \tau} \right),
\\
\psi_2 &= b_2^0 \left( \sin^2 \!\phi \, e^{-i\omega_+ \tau}
+ \cos^2 \!\phi \, e^{-i\omega_- \tau} \right).
\end{align*}
Thus the populations in states $|1\rangle$, $|2\rangle$ in consistent with Eqs. (\ref{subs}) are given by
\begin{gather*}
|a_1(-1, p_y{+}1)|^2 = |a_2^0(0, p_y)|^2
\frac{\alpha^2}{D_\text{eff}^2} \sin^2 \! D_\text{eff}\tau,
\\
|a_2(0, p_y)|^2 = |a_2^0(0, p_y)|^2 \left( 1 -
\frac{2\alpha^2}{D_\text{eff}^2} \sin^2 \! D_\text{eff}\tau \right),
\end{gather*}
where $2D_\text{eff}=\sqrt{ (\delta + 2p_y + 2 + \alpha)^2 + 8\alpha^2 }$. Atoms in the zero velocity group remain in state $|2\rangle$ for the pulse duration
\begin{equation}\label{duration}
\tau = \frac{2\pi}{ \sqrt{(\delta+2+\alpha)^2 + 8\alpha^2} },
\end{equation}
which corresponds to $|a_2(0, 0)|^2 = |a_2^0(0, 0)|^2$. One can easily see that in the limit of small $\alpha$ this result agrees with Eq.~(\ref{tau}).

\subsection{Full cooling process}

\begin{figure}
\includegraphics[width=8cm]{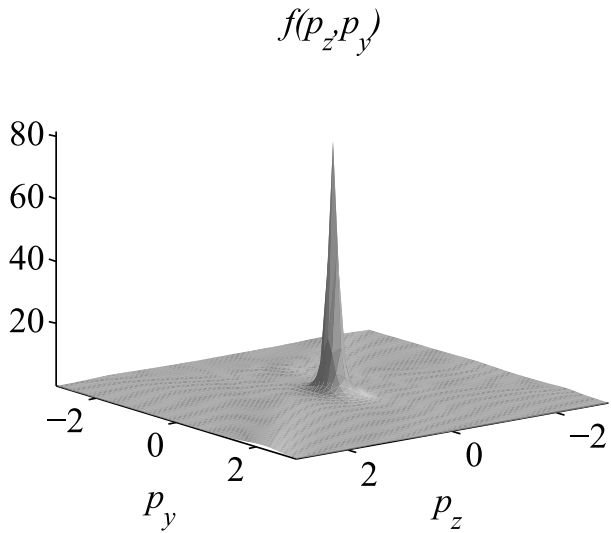}\caption{\label{distr}
The momentum distribution of atoms after applying 420 elementary cycles of 2D Raman cooling. Atoms have been piled up in a narrow peak near zero momentum with effective temperature equal to $0.01T_\text{rec}$. The height of the peak is about 80 times larger than that of the original distribution, the width (FWHM) is about 0.24 in units of $\hbar k$.}
\end{figure}

Full Raman cooling consists of varying Raman pulses, decreasing the detuning $\delta$ as in set (\ref{set}). After each six elementary cycles, the direction of $\pi$-polarized beam is alternated, so the next cycles excite another side of the velocity distribution. During cooling process, atoms are piled up in a narrow peak near zero momentum. As the number of Raman cycles increases, the number of atoms in the peak increases, so the momentum distribution becomes narrower, and the temperature of atoms decreases. The efficiency of growth increases as the Rabi frequency $\alpha$ of two-photon resonance increases. Numerical calculations show that deep Raman cooling is only achieved in the case of quite large $\alpha\tau$. Figure \ref{distr} shows the result after applying 420 elementary cycles of 2D Raman cooling with $\alpha=1$. The velocity spread of atoms is defined as $\sigma=(\text{FWHM})/\sqrt{8\ln 2}$ in units of the recoil velocity $v_\text{rec}$. It has been reduced from $3v_\text{rec}$ to $0.1v_\text{rec}$, corresponding to effective temperature $T_\text{eff}=0.01T_\text{rec}$.

Consider the 2D Raman cooling under conditions of experiment \cite{Theuer1999} where the same atomic and field configurations were realized in metastable Ne beam. The $\sigma_+$- and $\sigma_-$-polarized waves were provided by lasers with $\lambda_{Pump}=588 \text{ nm}$, while laser with $\lambda_{Stokes}=617\text{ nm}$ generated the $\pi$-polarized wave. Hence the final effective temperature of cooled atoms is given by
\begin{equation*}
T_\text{eff}\approx 0.01\frac{\hbar^2}{2 k_B M}(k_{Pump}^2 + k_{Stokes}^2)
= 26\text{ nK},
\end{equation*}
where $k_B$ is the Boltzmann constant, $M$ is the Ne atomic mass. The duration of only Raman pulses in our scheme is about 400~$\omega_R^{-1}$, where $\omega_R^{-1}$ for metastable Ne equals 6~$\mu$s. Full Raman cooling in turn takes a few milliseconds. The number $N_{\rm decay}$ of atoms spontaneously decaying from the upper state to a ground state during a Raman pulse is given by
\begin{equation*}
\dot N_{\rm decay} = \gamma N_4,
\quad
N_{\rm decay} \approx \frac{\gamma \tau}{\omega_R} N \frac{g^2}{\Delta^2}
= N \frac{\gamma}{\omega_R \Delta} \alpha\tau,
\end{equation*}
where $N$ is the number of atoms, $\gamma$ is the decay rate; the number $N_4 \approx N g^2/\Delta^2$ of atoms in the upper state is obtained from Eq. (\ref{appr}). Decay from the upper state can be neglected when $N_{\rm decay} \ll N_{1,3}$. Using Eqs. (\ref{a1}) and (\ref{a3}) one gets the inequality
\begin{equation}
N_{\rm decay} \ll N ( \alpha\tau )^2,
\end{equation}
which can be written in two alternative forms:
\begin{equation}
\alpha\tau \gg \frac{\gamma}{\omega_R \Delta},
\qquad
\frac{G^2}{\gamma^2} \gg \frac{\omega_R}{\gamma},
\end{equation}
where in the latter form it was assumed that $\tau \sim 1$ and $\alpha=G^2/\omega_R^2 \Delta$ (see Eq.~(\ref{scaling})). Thus by increasing either the detuning related to the upper state or the pulse intensities we can avoid the effects of spontaneous emission during the Raman pulses.

\section{Conclusions}

We have demonstrated that one can obtain very narrow velocity distributions ($\sigma\lesssim 0.1v_\text{rec}$) simultaneously in two directions with our straightforward scheme. The tripod approach presented here sets a further restriction on the available atomic structure, but this is not necessarily a major problem. Appropriate optically coupled $J=1$ and $J=0$ states are available \textit{e.g.} in Rb and Ne$^*$. In addition to 3D cooling by thermalization, the approach can be used to cool atoms in an 1D optical lattice, where the sample is ``sliced'' into 2D pancake-like structures. This approach is used for investigating optical frequency standards with alkaline-earth atoms~\cite{Boyd2007}. The simplicity of our approach, coupled with the expected efficiency, opens new possibilities for all-optical studies of ultracold atoms.

\section{Acknowledgments}
This research was supported by the Ministry of Education and Science of the Russian Federation, grants RNP 2.1.1/2166 and RNP 2.1.1/2532, by the Russian Foundation for Basic Research, grant RFBR 09-02-00223a, by the Academy of Finland, grant 133682, and by the Magnus Ehrnrooth
Foundation.

\end{document}